\documentclass{pnastwo}

\url{www.pnas.org/cgi/doi/10.1073/pnas.0709640104}
\copyrightyear{2008}
\issuedate{Issue Date}
\volume{Volume}
\issuenumber{Issue Number}

\usepackage{epsfig}
\usepackage{color}
\usepackage{tikz}
\usepackage{amssymb}
\usetikzlibrary{shapes}
\usepackage{amssymb}
\usepackage{bm}

\usepackage{hyperref}

\definecolor{red}{rgb}{1,0,0}
\definecolor{blue}{rgb}{0,0,1}
\definecolor{black}{rgb}{0,0,0}

\renewcommand{\be}{\begin{equation}}
\renewcommand{\ee}{\end{equation}}
\newcommand{\ba}{\begin{eqnarray}}
\newcommand{\ea}{\end{eqnarray}}

\newcommand{\eq}[1]{\begin{align}#1\end{align}}

\def\half{\mbox{$\frac{1}{2}$}}

\def\tr{\mbox{tr}}

\renewcommand{\p}{\partial}

\newcommand{\dz}{\delta z}
\newcommand{\PP}{\mathbb{P}}
\newcommand{\Gb}{\bm{G}}

\usepackage{amsmath}
\newlength{\arrow}
\begin{document}
\title{The force distribution affects vibrational properties in hard sphere glasses}

\author{E. DeGiuli\affil{1}{New York University, Center for Soft Matter Research, 4 Washington Place, New York, NY, 10003, USA}, E. Lerner\affil{1}{}, C. Brito\affil{2}{Instituto de Fisica da UFRGS, 3308-7286 Av. Bento Gon\c{c}alves, Porto Alegre, RS, Brasil}, M. Wyart\affil{1}{}}

\contributor{Submitted to Proceedings of the National Academy of Sciences
of the United States of America}

\significancetext{ How a liquid becomes rigid at the glass transition is a central problem in condensed matter physics. In many scenarios of the glass transition, liquids go through a critical temperature below which minima of free energy appear. However even in the simplest  glass -- hard spheres-- what confers mechanical stability at large density is highly debated.  In this work we show that to understand quantitatively stability at a microscopic level, the presence of weakly interacting pairs of particles must be included. This approach allows us to predict various non-trivial scaling behavior of the elasticity and vibrational properties of colloidal glasses, that can be tested experimentally. It also gives a spatial interpretation to recent calculations in infinite dimensions using methods widely used in glassy systems.}

\maketitle
\begin{article} 
\begin{abstract}

We study theoretically and numerically the elastic properties of hard sphere glasses, and provide a real-space description of their mechanical stability. In contrast to repulsive particles at zero-temperature, we argue that the presence of certain pairs of particles interacting with a small force $f$ soften elastic properties. This softening affects the exponents characterizing elasticity at high pressure, leading to experimentally testable predictions. Denoting $\PP(f)\sim f^{\theta_e}$ the force distribution of such pairs and $\phi_c$ the packing fraction at which pressure diverges, we predict that (i) the density of states has a low-frequency peak at a scale $\omega^*$, rising up to it as $D(\omega) \sim \omega^{2+a}$, and decaying above $\omega^*$ as $D(\omega)\sim \omega^{-a}$ where $a=(1-\theta_e)/(3+\theta_e)$ and $\omega$ is the frequency, (ii) shear modulus and mean-squared displacement are inversely proportional with $\langle \delta R^2\rangle\sim1/\mu\sim (\phi_c-\phi)^{\kappa} $ where $\kappa=2-2/(3+\theta_e)$, and (iii) continuum elasticity breaks down on a scale $\ell_c \sim1/\sqrt{\delta z}\sim (\phi_c-\phi)^{-b}$ where $b=(1+\theta_e)/(6+2\theta_e)$ and $\delta z=z-2d$, where $z$ is the coordination and $d$ the spatial dimension. We numerically test (i) and provide data supporting that $\theta_e\approx 0.41$ in our bi-disperse system, independently of system preparation in two and three dimensions, leading to  $\kappa\approx1.41$, $a \approx 0.17$, and $b\approx 0.21$. Our results for the mean-square displacement are consistent with a recent exact replica computation for $d=\infty$, whereas some observations differ, as rationalized by the present approach. 



\end{abstract}


The emergence of rigidity near the glass transition is a fundamental and highly debated topic in condensed matter, and is perhaps most surprising in hard sphere glasses where rigidity is purely entropic in nature. The rapid growth of relaxation time around a packing fraction $\phi_g\approx 0.58$ suggests that meta-stable states have appeared in the free energy landscape, and that activation above barriers is required for the system to flow \cite{Goldstein69}. This scenario is presumably what Mode Coupling Theory captures \cite{Kirkpatrick89,Berthier11b}, can be rationalized via density functional theory \cite{Singh85} and via the replica method \cite{Parisi10}. Recently  a real-space description of mechanical stability and elasticity in hard sphere glasses has been proposed \cite{Brito06,Brito09}, which is most easily tested at large pressure, deep in the glass phase. It is based on two results. First, in elastic networks and athermal packings of soft spheres \cite{Liu10,Hecke10,Wyart05b}, mechanical stability is controlled by the mean number of contacts per particle, or coordination $z$ (as already discussed by Maxwell \cite{Maxwell64}), and the applied compressive strain $e$ \cite{Wyart05b}. As one may intuitively expect, increasing coordination is stabilizing, whereas increasing pressure at fixed coordination is destabilizing.  Second, within a long-lived metastable state the vibrational free energy of a hard sphere system can be approximated as a sum of local interaction terms between pairs of colliding particles, which are said to be ``in contact". On a time scale that contains many collisions, at high packing fraction the interaction follows approximately $V(h)\approx-k_B T \log h$  where $h$ is the time-averaged distance between two adjacent particles \cite{Brito06,Brito09}. This directly leads to an effective force law $f(h) \approx k_B T/h$ and allows one to map a hard sphere system near the random close packing $\phi_c$ to a zero-temperature elastic network. 
These two sets of results yield a stability constraint on the microscopic structure of hard sphere glasses, which in practice appears to lie very close to saturation \cite{Brito06,Brito09,Ikeda13}. Such {\it marginal stability} implies the  abundance of very soft elastic modes, as confirmed empirically \cite{Brito06,Brito09,Ghosh10,Chen10,Ikeda13,Kaya10,Mari09}, and fixes the scaling behavior of elasticity as jamming is approached \cite{Brito09}. In particular the particles' mean-squared displacement was predicted to follow $\langle \delta R^2\rangle \sim (\phi_c-\phi)^\kappa$ with $\kappa=1.5$ \cite{Brito09} instead of the naive $\kappa=2$, which would hold in a crystal: particles in the glass fluctuate much more than the size of their cage (defined as the typical distance between particles), due to the presence of collective soft modes.

Very recently a replica calculation \cite{Charbonneau13, Charbonneau14} predicted $\kappa=1.41574$ in infinite dimensions, close but different from the prediction of \cite{Brito06,Brito09}. At $\phi_c$ it also predicted for the force distribution $\PP(f)\sim f^{\theta_f}$ with $\theta_f=0.42311$ and for the gap distribution $g(h)\sim h^{-\gamma}$ with $\gamma=0.41269$. Some of these latter results are consistent, and some differ, from an earlier analysis based on the stability of jammed packings (at $\phi_c$) toward changes of their network of contacts \cite{Wyart12,Lerner13}. In these works $\gamma$ was argued and shown numerically to be related to the force distribution exponents $\theta_e$ and $\theta_\ell$, characterizing respectively two kinds of contacts at low forces \cite{Lerner13} (see below). Here we propose a resolution of these issues: heterogeneity in contact strength was neglected in \cite{Brito06,Brito09}, but the prevalence of weak forces in hard-sphere systems corrects scaling exponents, and leads to the scaling relation $\kappa=2-2/(3+\theta_e)$ consistent with the result of \cite{Charbonneau13}, if $\theta_f = \theta_e$ in dimension $d=\infty$. 
 We compute the associated modification in the scaling of elastic properties as $\phi\rightarrow \phi_c$. Furthermore, we argue that some key properties of packing differ in finite and infinite dimensions, so that $\theta_f =\theta_\ell$ in $d=2, 3$ while $\theta_f = \theta_e$ in $d= \infty$.
 In general, our approach leads to a description of the structure of packings in terms of four exponents related by three scaling relations. 
 
This work is organized as follows: in the Section {\it Elastic Networks}, we present a variational argument for the density of vibrational modes in weakly-coordinated networks with stiffness heterogeneity. We also use scaling arguments to compute the shear modulus and the mean-squared displacement. In Section {\it Effective Medium Theory}, we confirm these predictions using a standard mean-field approximation, and furthermore predict the length scale below which continuum elasticity breaks down in such systems. In the Section {\it Hard Spheres} we show how these results apply to colloidal glasses, and discuss the subtle issue associated with the existence of two kind of contacts at low forces in sphere packings. We also present numerical results supporting our views. In the last two sections, we compare our results with replica calculations, and discuss prospects for experimental tests in colloidal systems.

\begin{figure}[t!] 
\begin{tikzpicture}[scale=1]
\node[above right] {\includegraphics[viewport=22 73 450 280,width=0.49\textwidth,clip]{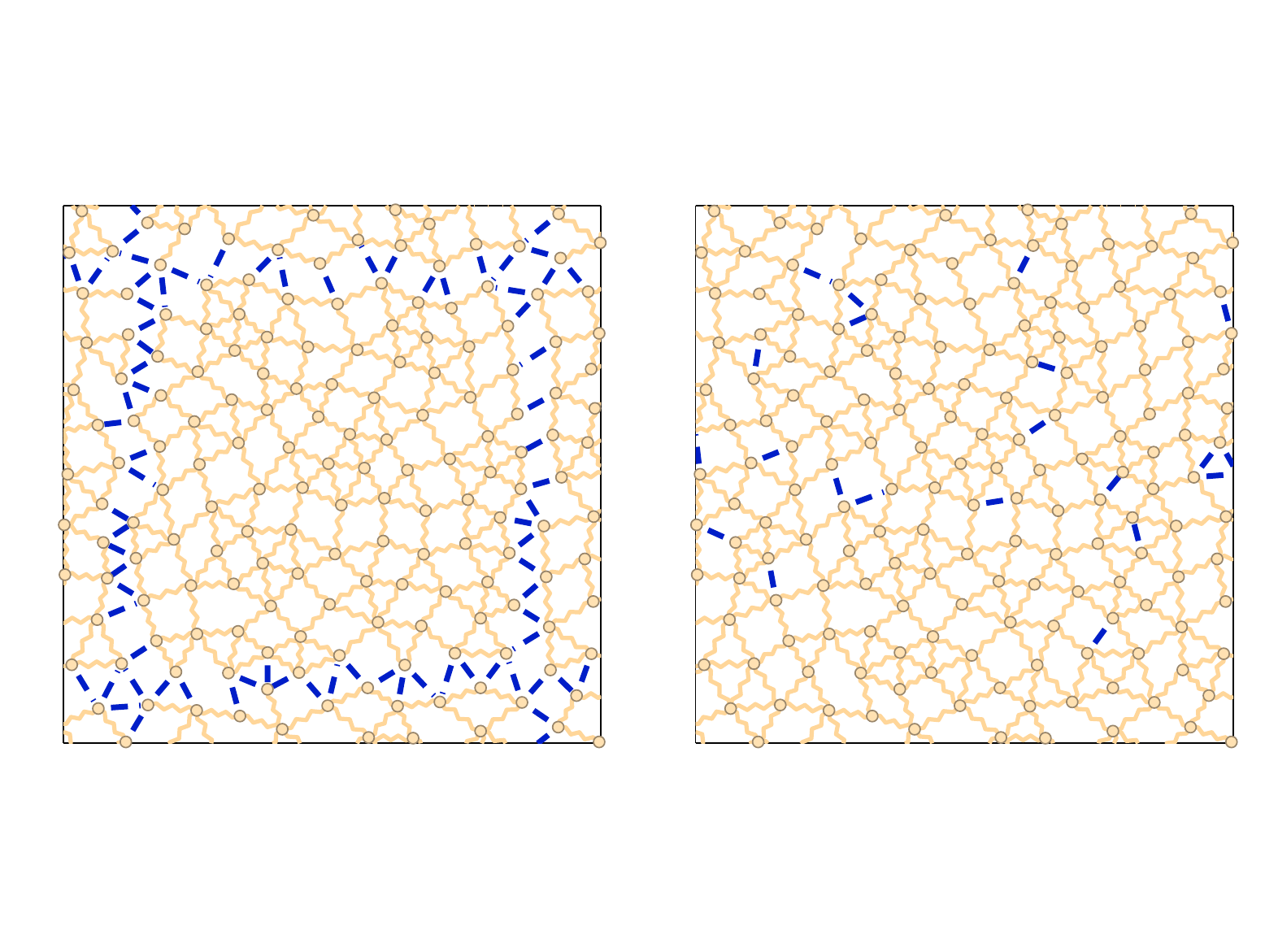}}; 
\draw[fill=white] (0.2,3.3) rectangle (0.8,3.9);
\draw(0.81,3.6) node[left] {(a)};
\draw[fill=white] (4.8,3.3) rectangle (5.4,3.9);
\draw(5.41,3.6) node[left] {(b)};
\end{tikzpicture}
\caption{\label{fig:cut} Illustrative diagram of cutting argument, showing cut bonds in blue (thick lines). (a) Bonds are cut around blocks of size $L\times L$, a useful procedure when $\alpha>0$; (b) When $\alpha<0$, the variational argument is improved by cutting instead the fraction $q$ of weakest bonds. }
\end{figure}
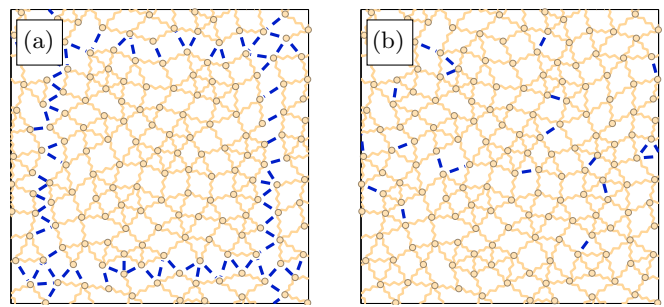
\section{Elastic Networks:}
We consider an elastic network of $N$ points of mass $m$, connected by $N_c$ springs, of  coordination $z=2N_c/N$, in spatial dimension $d$. 
The quadratic expansion of the elastic energy for an imposed displacement field $|\delta R\rangle$  follows \cite{Landau60,Alexander98}:
\ba
\label{1}
\delta E\equiv \frac{1}{2} \langle \delta R|{\cal M}|\delta R\rangle= 
\frac{1}{2} \sum_{\beta} k_{\beta} \; \delta R^\parallel_\beta{}^2  -\frac{f_{\beta}}{r_{\beta}} \; \delta R^\perp_\beta{}^2
\ea
where the sum is over springs $\beta$.  
Here $r_\beta$, $k_\beta$, and $f_\beta$ are the spring length, stiffness, and force (chosen positive for a repulsive interaction), and $\delta R^\parallel_\beta$ and $\delta R^\perp_\beta$ are, respectively, the magnitude of displacements parallel and perpendicular to the spring $\beta$, i.e., $\delta R^\parallel_\beta = (\delta {\vec R}_i-\delta {\vec R}_j)\cdot {\vec n}_{\beta}$, and  $\delta R^\perp_\beta = |\delta {\vec R}_i-\delta {\vec R}_j - {\vec n}_{\beta} \delta R^\parallel_\beta|$, where ${\vec n}_{\beta}$ is a unit vector along the spring $\beta$. 


We assume that the $r_\beta$ are narrowly distributed about their mean $\langle r_\beta\rangle=\sigma$ which defines our unit length, and introduce $k_c\equiv \langle k_\beta \rangle$ and $\omega_c=\sqrt{k_c/m}$. Eq.(\ref{1}) defines the stiffness matrix ${\cal M}$, which is positive definite in a stable configuration. The eigenvalues of ${\cal M}$ are $\lambda=m\omega^2$, where the $\omega$'s are the frequencies of vibrational modes, of density $D(\omega)$.  

\subsection{ Variational argument} 

First we consider the springs at rest length, so that all $f_{\beta}=0$ and only the parallel term in \eqref{1} is present.  Let $\dz \equiv z - z_c$ with $z_c=2d$. As pointed out by Maxwell, if $Nd>N_c$ (or equivalently $\dz < 0$) it is clear from Eq.(\ref{1}) that there are at least $Nd-N_c$ displacement fields with no restoring force ($\delta E=0$), the so-called floppy modes.  They are solutions to the set of linear equation $\delta R^\parallel_\beta=0\ \ \forall \beta$. We assume that the shape of the stiffness distribution $\PP(k)$ is independent of $z$, and wish to compute the scaling properties of $D(\omega)$ as $\dz \to 0^+$. Our strategy is to build trial modes, which are orthonormal displacement fields with small energy. Using the fact that ${\cal M}$ is positive definite then allows one to bound from below the number of eigenvalues below some threshold, leading to a lower bound on $D(\omega)$. This strategy was used in \cite{Wyart05,Wyart05b}, where trial modes were constructed from the floppy modes that appear by cutting the system into compact regions of size $L$, as shown in Fig. \ref{fig:cut}a. This requires cutting a fraction $q \sim 1/L$ of bonds. For a system at $\dz=0$, the density of induced floppy modes per particle is simply $(dN-(1-q)N_c)/(dN) = q \sim 1/L$. These modes can be distorted to lead to trial modes of frequency $\omega(L) \sim \omega_c/L \sim \omega_c q$ in the original, uncut system \cite{Wyart05}. Since the density of states is the density of modes per unit frequency, one gets $D(\omega) \gtrsim q(\omega)/\omega \sim \omega^0/\omega_c$, implying that the vibrational spectrum does not vanish at zero frequency at the Maxwell bound. If $\dz > 0$, then when a fraction $q$ of bonds are cut, the density of induced floppy modes is $q-\dz/z_c$. This leads to a cut-off frequency $\omega^* \sim \omega_c \; \dz$, such that $D(\omega) \gtrsim 1/\omega_c$ above $\omega^*$, as observed numerically \cite{Ohern03,Silbert05,Wyart05b,Liu10,Hecke10}. 

We now show that if the distribution of stiffnesses is broad enough, then the above bound is not saturated. In this case, we can improve the variational argument by creating a different set of trial modes, illustrated in Fig. \ref{fig:cut}b; we cut a fraction $q$ of the weakest links, and use the density $q-\dz/z_c$ of induced floppy modes. We then make the key assumption that these floppy modes do not decay appreciably with distance from the broken bonds, but extend in the entire system, displacing particles by some characteristic amplitude. On the one hand, this assumption is supported by the proof that in an isostatic system, the response to a local strain does not decay as a power-law of distance \cite{Wyart05b}, unlike what occurs in a normal (well-connected) elastic medium. On the other hand, this argument does not exclude the possibility that floppy modes have a very large amplitude just where the contacts were cut, and then a small background displacement not decaying with distance. We shall see below that for hard spheres, our assumption only holds for a fraction of the contacts at low-force. 

By definition, the displacements of floppy modes are strictly perpendicular to bonds, except at the broken bonds themselves. In particular, if we cut the bond $\beta$,  $\delta R^\parallel_\gamma = 0$ for all $\gamma \neq \beta$. Our assumption that floppy modes are extended means that $\delta R^\parallel_\beta \sim \langle |\delta{\vec  R_i}| \rangle \sim 1/\sqrt{N}$ where the average is made on all particles $i$, and the last equation reflects normalization. 


We assume that the distribution of stiffnesses follows 
$\PP(k)\sim k^\alpha/k_c^{\alpha+1}$ at low stiffnesses, where $\alpha>-1$. 
Let $\dz \equiv z-z_c > 0$. The fraction $q$ of weakest extended bonds have a 
characteristic stiffness $k_0$ with $\int_0^{k_0} \PP(k) dk = q$, 
leading to $k_0\sim k_c \; q^{1/(1+\alpha)}$. A density $q-\dz/z_c$ of modes in the system are floppy. In the original system, these modes stretch or compress the fraction $q$ of weak springs of characteristic stiffness $k_0$, and thus have a finite energy of order $E \propto \sum_{\beta} k_\beta \delta R_\beta^\parallel{}^2 \sim qN k_0 1/N = q k_0$, 
leading to a characteristic frequency:
\be
\label{3}
\omega(q)\propto \sqrt{E/m}\sim \omega_c \; q^{(2+\alpha)/(2+2\alpha)}
\ee
The variational inequality implies $D(\omega) \gtrsim (q-\dz/z_c)/\omega$. This argument can be applied with any $q \ll 1$ such that $q > \dz/z_c$, implying that $\omega \gtrsim \omega_c \; (\dz/z_c)^{(2+\alpha)/(2+2\alpha)}$. It is convenient to let $q=r \dz$ with $r>1/z_c$. Then
\be
\label{4}
D(\omega)\gtrsim \frac{(r-1/z_c)}{\omega_c} \left(\frac{\omega}{\omega_c}\right)^{\!\alpha/(2+\alpha)}
\ee
These are our central results: at the Maxwell threshold ($z=z_c$), when weak interactions are abundant ($\alpha<0$), the density of states must diverge at zero frequency, with a non-trivial exponent. When the coordination is larger ($z>z_c$), the scaling for $D(\omega)$, Eq. \eqref{4}, holds above the characteristic frequency:
\be
\label{w*}
\omega^*\sim \omega_c \;(\dz/z_c)^{(2+\alpha)/(2+2\alpha)}.
\ee
For $\alpha>0$ the new bound is not useful and the previous argument of \cite{Wyart05} applies. Note that in all cases we consider $q \ll 1$ so that $\omega \ll \omega_c$. 
Assuming harmonic dynamics and Eq.(\ref{4}), one obtains a bound for the particles' mean-squared displacement $\langle\delta R^2\rangle$:
\be
\label{6}
\frac{k_c\langle  \delta R^2\rangle}{k_B T}\!=  \omega_c^2\!\int\!\frac{D(\omega)}{ \omega^2} d\omega>\omega_c^2\!\int_{\omega>\omega^*}\!\!\!\frac{D(\omega)}{ \omega^2} d\omega \gtrsim\!\left(\!\frac{\omega^*}{\omega_c}\!\right)^{\frac{-2}{2+\alpha}}
\ee

To estimate the shear modulus, we cut a fraction $q=2 \dz/z_c$ of the weakest links, so that the system is now floppy with a density of floppy modes $\dz/z_c$, and no elasticity. It was shown \cite{Wyart08,Lerner12} that under an applied shear of strain $\epsilon$, the relative displacement of particles (of order of the non-affine displacement) is of order $\epsilon/\sqrt{\dz}$, as observed numerically  \cite{Wyart08,Ellenbroek09,Ellenbroek06}. In the uncut system, this deformation has energy $\delta E \sim q k_0(q) (\epsilon/\sqrt{\delta z})^2$, leading to a shear modulus:
 \be
 \label{5}
 \mu\sim k_0\sim k_c \delta z^{1/(1+\alpha)}
 \ee

\begin{figure}[t] 
\centering
\begin{tikzpicture}[scale=0.96]
\clip (-0.5,0) rectangle (9,3.7);
\node[above right]at (-0.5,0){\includegraphics[width=0.49\textwidth]{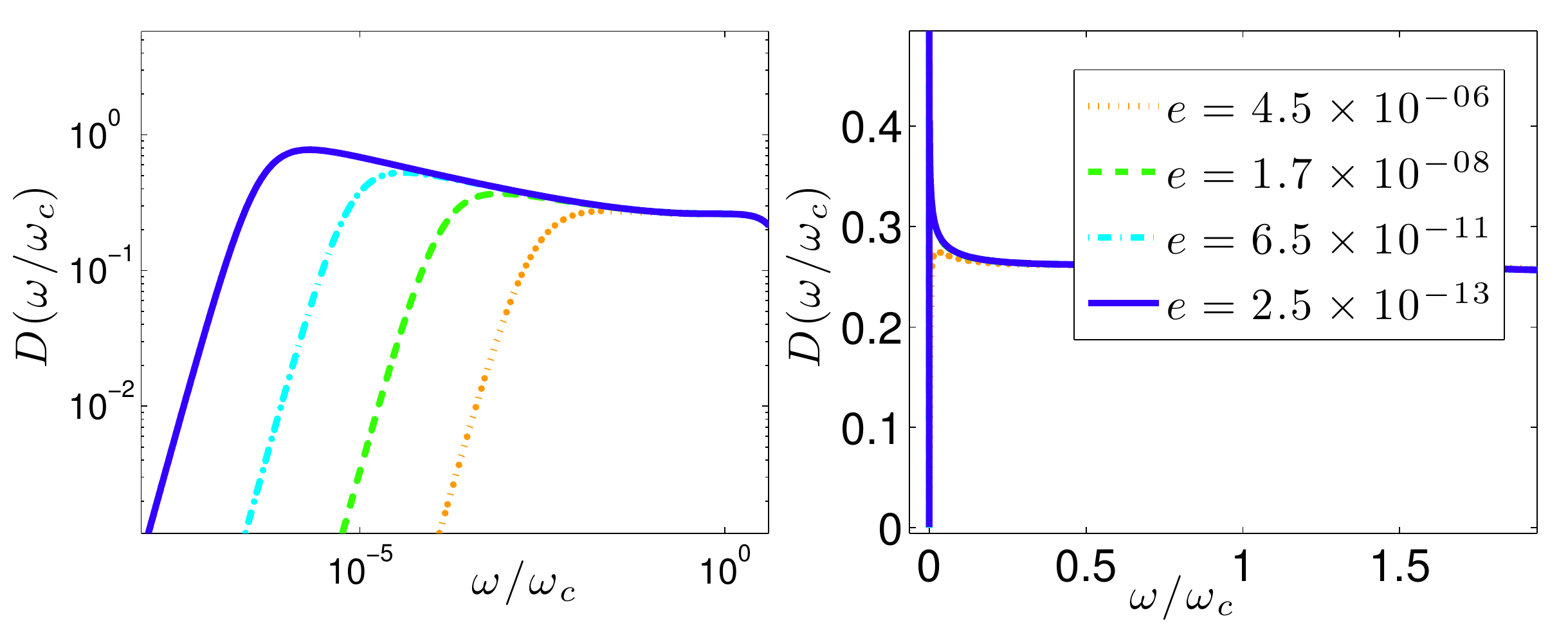}};
\caption{\label{fig:Domega1} Effective medium theory prediction for the density of states $D(\omega)$ for a marginally-stable material in $d=2$ with $\alpha = -0.30$, at indicated values of $e=\phi_c-\phi$, in (left) log-log axes, and (right) linear axes. As discussed in the text, this corresponds to a hard-disk glass with $\theta_e=0.41$. The peak appears at the frequency scale $\omega^*~\sim~\omega_c \sqrt{e}$.}
\end{tikzpicture}
\end{figure}

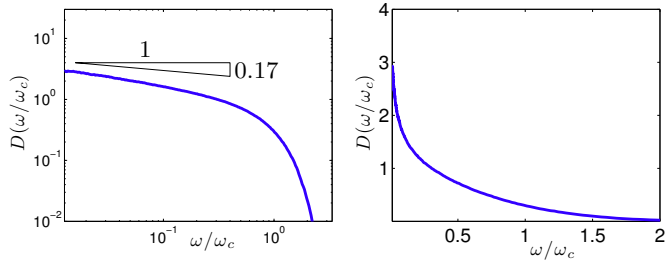
\begin{figure}[t] 
\centering
\begin{tikzpicture}[scale=0.96]
\clip (-0.5,0.05) rectangle (9,3.7);
\node[above right]at (-0.5,0){\includegraphics[width=0.49\textwidth]{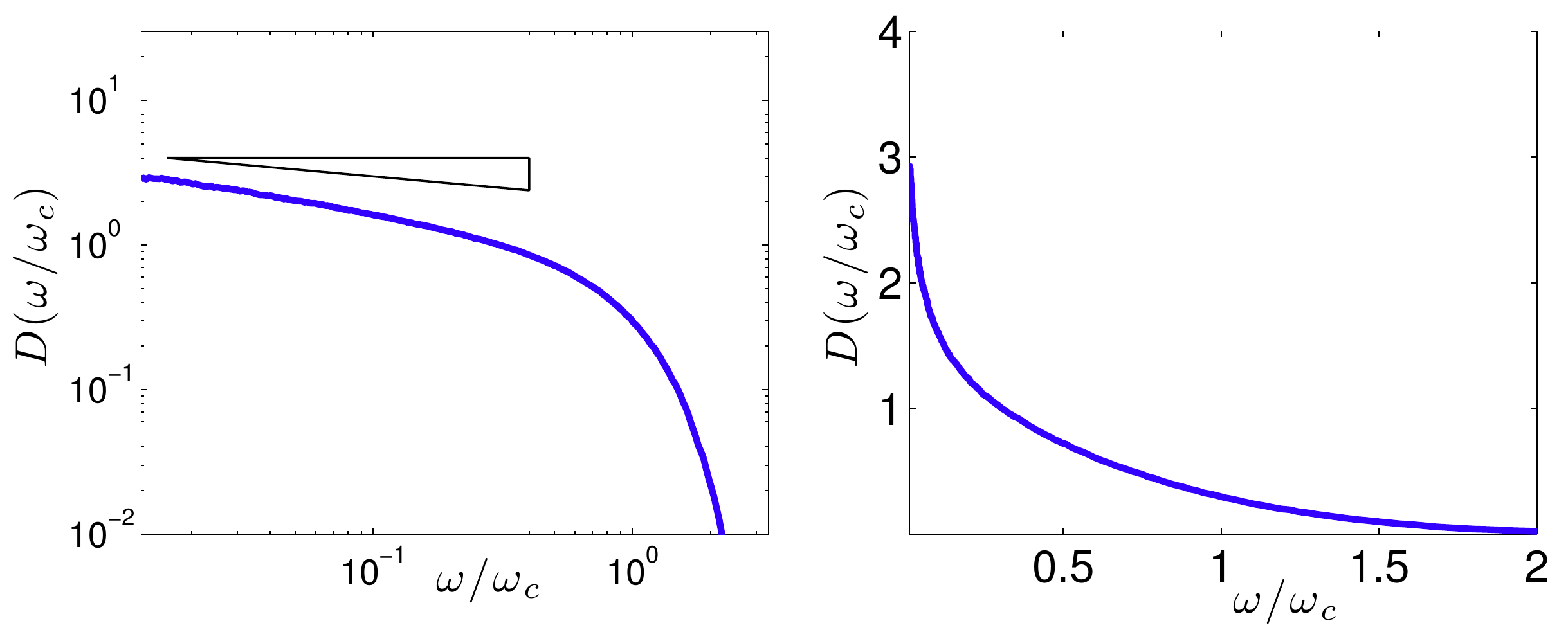}};
\draw(1.55,2.83) node[above] {$1$};
\draw(2.65,2.77) node[right] {$0.17$};
\end{tikzpicture}
\caption{\label{fig:Domega3} Numerical density of states $D(\omega)$ for a hard-sphere glass in $d=2$, at pressure $p=10^{12}$, in (left) log-log axes, and (right) linear axes. The triangle has the predicted slope $-0.17$, assuming $\theta_e=0.41$, as discussed in the main text. The characteristic frequency $\omega^*$ is expected to be $\sim 10^{-6}\omega_c$, outside the accessible numerical range at this pressure. }
\end{figure}

\subsection{Role of pre-stress} The presence of a compressive force in the bonds reduces the modes' frequency, as implied by Eq.(\ref{1}), and can lead to an elastic instability. It was argued and checked numerically in  \cite{Wyart05b} that the strongly scattered modes that appear above $\omega^*$ have large relative displacements, of order of the displacement itself: $|\delta R^\perp_\beta | \sim |\delta {\vec R}_i|$. Following Eq.(\ref{1}) this implies that some soft modes will be shifted to a frequency $\omega_0$ satisfying $\delta E\equiv m\omega_0^2= m\omega^*{}^2- A  f_c$, where $f_c$ is the characteristic compressive force and $A$ a numerical constant. Stability requires $\omega_0>0$, implying 
\be
\label{w0}
\omega^* \gtrsim \omega_c \sqrt{e},
\ee
where we have defined the contact strain $e\equiv f_c/k_c$. Using Eq.(\ref{w*}) this becomes $\delta z \gtrsim e^{(1+\alpha)/(2+\alpha)}$, 
 extending the previous result $\delta z\gtrsim\sqrt e$  \cite{Wyart05b} to the case $\alpha<0$. In packings of particles, $e\propto |\phi-\phi_c|$ and the latter bound was argued to be saturated, based on dynamical considerations \cite{Wyart05b,Brito06,Brito09}.


\section{Effective Medium:} All the above predictions can be derived and extended with effective medium theory (EMT), a mean-field approximation that treats disorder in a self-consistent way \cite{Garboczi85,Schirmacher07,  Webman81,Wyart10a,Mao10,DeGiuli14,Sheinman12}. 
EMT has been shown to give quantitatively correct values for scaling exponents related to the vibrational spectrum and heat transport properties of frictionless packings \cite{Wyart10a,DeGiuli14}. In EMT, a random elastic network, such as depicted in Figure \ref{fig:cut}, is modeled by a regular lattice with effective frequency-dependent spring constants. Here we follow the EMT developed in \cite{DeGiuli14} which includes the effect of forces in Eq.(\ref{1}). In \cite{DeGiuli14} the randomness in the interaction between two nodes was limited to the presence or absence of a spring; when a spring was present, its stiffness was always identical. Here we relax this assumption and allow a full distribution of stiffnesses, behaving as $\PP(k) \sim k^\alpha$ for small $k$, and allow a distribution of contact forces, $\PP(f) \sim f^{\theta_f}$ at small $f$. Details of the EMT are presented in Supplementary Information (SI)\cite{HardSphereSI}. 

The EMT confirms that when $\alpha >0$, previous results of \cite{Wyart05b,DeGiuli14} are obtained. When $\alpha < 0$, in addition to confirming the scaling results presented above, EMT gives the form of the complex shear modulus and density of states when $\dz$ is small, and can be used to extract other vibrational and heat transport properties. In general, two frequency scales are predicted, as in the variational argument: $\omega^*$ and $\omega_0 = \omega^* \sqrt{1-e/e_c}$, where $e_c\sim \dz ^{(2+\alpha)/(1+\alpha)}$ is the contact strain at elastic instability \cite{DeGiuli14}. For a marginally stable material, $e\approx e_c$ and therefore $\omega_0/\omega^* \ll 1$. Above its peak at $\omega^*$, EMT predicts that $D(\omega)$ decays as $D(\omega/\omega_c) \sim (\omega/\omega_c)^{\alpha/(2+\alpha)}$,  in agreement with Eq.~(\ref{4}), with a logarithmic correction in $d=2$. Between $\omega_0$ and $\omega^*$, EMT predicts 
\eq{ \label{Domega2}
D(\omega) \sim \frac{1}{\omega_c} \left(\frac{\omega}{\omega_c}\right)^{1+2/(\alpha+2)} \left(\frac{\omega^*}{\omega_c}\right)^{-4/(\alpha+2)}
}
Numerical solution of the leading-order EMT equation for a marginally stable material in $d=2$ gives the result shown in Figure \ref{fig:Domega1}, where we have taken $\alpha = -0.30$. The visible curvature is due to logarithmic corrections, which are only present in $d=2$.

Regarding the shear modulus, EMT 
confirms the scaling Eq.(\ref{5}), and in addition we find the dependence on $e/e_c$. At fixed $\dz$, we find that $\mu$ drops by a finite factor at elastic instability, relative to its unstressed value. Finally, EMT predicts that modes at $\omega^*$ have 
a scattering length $\ell_c \sim \dz^{-1/2}$, also characterizing the response to a point force \cite{Lerner13a}.


\section{Hard spheres:}
The above results on elastic networks can be applied to the free energy of hard spheres within a metastable state, and near maximum packing at $\phi_c$. To do so, we consider a mesoscopic time scale $\tau$, much larger than the typical interval between collisions, $\tau_C$, and define a `contact' network by those particles that collide on the time scale $\tau$ \cite{Wyart05b,Brito06,Brito09}. Using the fact that the contact network at $\phi_c$ is isostatic, one can show that the Helmholtz free energy of the metastable state is well approximated by a sum of two-body effective potentials, which follow 
\eq{ \label{V}
V(h) \approx -k_B T \log h, 
}
where $h$ is the time-averaged gap between `contacting' particles. Hence in link $\beta$ the force $f_\beta \approx k_B T/h_\beta$, and the stiffness $k_\beta \approx k_B T/h^2_\beta$. It was checked previously in simulations that this effective potential is very closely followed near $\phi_c$, and in particular deviations are less than 5\% within the glass phase \cite{Brito06,Brito09}. We therefore assume that the effective potential is fixed and independent of $z$.

The distribution of contact forces at $\phi_c$ is known to follow $\PP(f)\sim f^{\theta_f}$ at small $f$, with $\theta_f \approx 0.2$ \cite{Charbonneau12,Lerner13}. This directly yields a diverging distribution of stiffnesses: $\PP(k) = \PP(f) df/dk \sim k^\alpha$, with $\alpha=-(1-\theta_f)/2 < 0$. Hence there are indeed very many contacts with a weak stiffness. However, to apply our earlier results, we have also assumed in the variational argument that each opened weak link induces an extended mode that does not decay appreciably with distance. This condition leads to a subtlety in the exponent $\alpha$. 

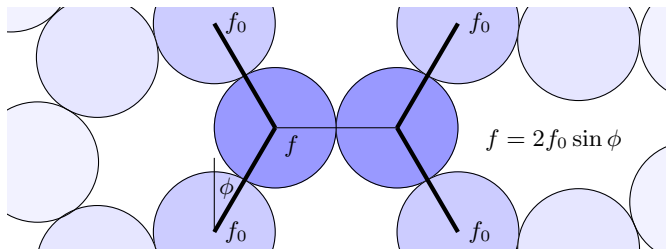
\begin{figure}[t!] 
\centering
\begin{tikzpicture}[scale=0.8]
\clip (-5.4,-2) rectangle (5.4,2);
\filldraw[color=blue!40!white](1,0) circle (1);
\draw(1,0) circle (1);
\filldraw[color=blue!40!white](-1,0) circle (1);
\draw(-1,0) circle (1);
\filldraw[color=blue!20!white](2,-1.73205) circle (1);
\draw(2,-1.73205) circle (1);
\filldraw[color=blue!20!white](2,+1.73205) circle (1);
\draw(2,+1.73205) circle (1);
\filldraw[color=blue!20!white](-2,-1.73205) circle (1);
\draw(-2,-1.73205) circle (1);
\filldraw[color=blue!20!white](-2,+1.73205) circle (1);
\draw(-2,+1.73205) circle (1);
\filldraw[color=blue!10!white](2+2*0.99,+1.73205-2*0.14) circle (1);
\filldraw[color=blue!10!white](2+2*0.99,-1.73205-2*0.14) circle (1);
\filldraw[color=blue!5!white](2+2*0.99+2*0.92,-1.73205-2*0.14+2*0.39) circle (1);
\draw(2+2*0.99,+1.73205-2*0.14) circle (1);
\filldraw[color=blue!5!white](2+2*0.99+2,+1.73205-2*0.14) circle (1);
\draw(2+2*0.99+2,+1.73205-2*0.14) circle (1);
\draw(2+2*0.99,-1.73205-2*0.14) circle (1);
\draw(2+2*0.99+2*0.92,-1.73205-2*0.14+2*0.39) circle (1);
\filldraw[color=blue!10!white](-2-2*0.96,+1.73205-2*0.28) circle (1);
\filldraw[color=blue!5!white](-2-2*0.96-2*0.71,+1.73205-2*0.28+2*0.71) circle (1);
\draw(-2-2*0.96-2*0.71,+1.73205-2*0.28+2*0.71) circle (1);
\filldraw[color=blue!10!white](-2-2*0.96,-1.73205-2*0.28) circle (1);
\filldraw[color=blue!5!white](-4.9,-0.57) circle (1);
\draw(-2-2*0.96,+1.73205-2*0.28) circle (1);
\draw(-2-2*0.96,-1.73205-2*0.28) circle (1);
\draw(-4.9,-0.57) circle (1);
\draw(-1,0) -- (1,0);
\draw(-1,-0.3) node[right] {$f$};
\draw(2.3,-0.2) node[right] {$f=2f_0 \sin\phi$};
\draw[ultra thick](1,0) -- (2,1.73205) node[right]{$f_0$};
\draw[ultra thick](-1,0) -- (-2,1.73205) node[right]{$f_0$};
\draw[ultra thick](1,0) -- (2,-1.73205) node[right]{$f_0$};
\draw[ultra thick](-1,0) -- (-2,-1.73205) node[right]{$f_0$};
\draw(-2,-1.73205) -- (-2,-0.5);
\draw(-1.8,-1.0) node{$\phi$};
\end{tikzpicture}
\caption{\label{figlocal} Illustration of a local configuration of particles that gives rise to small displacements when opening the central horizontal contact. Line thickness represents, schematically, force magnitude in the central region. Even if the force $f_0$ in the surrounding contacts is on the order of the mean force, $f_0 \sim \langle f \rangle$, the force in the horizontal contact can be small if the angle $\phi$ is small, and displacements resulting from opening that contact will be of order $\delta R \sim \sin(\phi)$.}
\end{figure}

In \cite{Lerner13} it was observed that when contacts are opened from hard sphere packings at $\phi_c$, there are in addition to the `extended' modes discussed above, also `localized' modes: deformations that decay on the scale of a few grains. Such `localized' modes occur because of local correlations in the structure, as illustrated in Figure \ref{figlocal}. In SI we show that the variational argument is not improved by including the localized contacts, and therefore we want to consider only the extended type. In $d=2$ and $d=3$, the distribution of localized contacts was observed to follow $f^{\theta_\ell}$ with $\theta_\ell \approx 0.17$, while that of the extended contacts follows $f^{\theta_e}$ with $\theta_e \approx 0.44$  \cite{Lerner13}. Since the localized contacts are more numerous, the distribution of forces follows $\PP(f)\sim f^{\theta_f}$ with $\theta_f=\theta_\ell$. However, only the extended contacts can be included in our theory, therefore we have $\alpha = -(1-\theta_e)/2$.

We can now present our results for hard spheres. Geometrically, the characteristic gap $h_c \sim \phi_c-\phi$, so that the characteristic force and stiffness are, respectively, $f_c \sim k_B T/h_c$ and $k_c \sim k_B T/h_c^2$. 
Stability requires the Hessian is positive-definite, and therefore following Eq.(\ref{w0}) that $\omega^*\gtrsim \omega_c (f_c/k_c)^{1/2} \sim (\phi_c-\phi)^{-1/2}$, a result identical to the previous approach \cite{Brito06,Brito09} neglecting stiffness heterogeneity. In \cite{Brito09,Ikeda13} this bound was observed to be saturated, and here we assume such marginal stability, $\omega^*/\omega_c \sim (\phi_c-\phi)^{1/2}$. From Eqs.(\ref{4},\ref{w*},\ref{6},\ref{5},\ref{Domega2}) we then deduce 
\ba
D(\omega)&\sim&\begin{cases} & \!\!\omega^{2+a} \ \ \ \ \ \ \ \ \ \ \ \ \  \ \  \ \hbox{for       } \;\;\;\; \;\ \ \ \ \ \omega < \omega^* \\
& \!\! \omega^{-a} \ \ \ \ \ \ \ \ \ \ \ \ \ \  \ \  \ \hbox{for       } \;\; \omega^* < \omega \ll \omega_c \label{6bis} \end{cases} \\
\label{7}
\langle \delta R^2\rangle&\sim&\frac{1}{\mu} \sim (\phi_c-\phi)^\kappa \\
\label{7bis}
\dz &\sim& (\phi_c-\phi)^{2b}
\ea
where
\eq{
a=\frac{1-\theta_e}{3+\theta_e}, \;\; b=\frac{1+\theta_e}{6+2\theta_e}, \;\; \kappa=\frac{4+2\theta_e}{3+\theta_e}. \label{exp}
}
Using that the pressure $p \sim f_c$, our prediction for $\dz(p)$ appears satisfied in recent simulations \cite{Charbonneau12} if it is assumed that the contact network corresponds to those particles closer than a characteristic gap $h^\dagger$ where $g(h)$ changes behavior (see SI). In SI, we argue that these results are not changed if the evolution of $\PP(k)$ with packing fraction is taken into account.

The new scaling relation (\ref{7}) relates two experimentally accessible quantities, $\langle \delta R^2\rangle$ and $\phi_c-\phi$, but through an exponent $\kappa$ that depends on $\theta_e$, which is not easily measurable. In \cite{Wyart12,Lerner13}, stability of jammed packings at $\phi_c$ was shown to relate the exponent $\gamma$ describing the distribution of gaps between particles, $g(h)\sim h^{-\gamma}$ \cite{Donev05a,Silbert06,Charbonneau12,Lerner13}, and the exponents $\theta_e$ and $\theta_\ell$. In particular, triggering one of these contact-opening excitations can lead to rewiring of the contact network. Stability of the system to extensive avalanches of rewiring was shown to imply  \cite{Wyart12,Lerner13} 
\eq{
\label{10}
\gamma & \geq \frac{1-\theta_\ell}{2} \\
\gamma & \geq \frac{1}{2+\theta_e} \label{11}
}
In \cite{Lerner13} it was observed that contact-opening excitations in packings are marginally stable, so that the the bounds (\ref{10}) and (\ref{11}) are satisfied with equality, with numerical values $\gamma\approx 0.4$, $\theta_\ell \approx 0.17$ and $\theta_e\approx 0.44$. Indeed assuming such marginal stability, it follows that $\theta_f=\theta_\ell<\theta_e$ and the exponent $\theta_e$ can be determined from $\theta_e = 2\theta_f/(1-\theta_f)\approx 0.41$, a value  consistent with the direct measurement $0.44$. 

Equations (\ref{7}), (\ref{10}), and (\ref{11}) lead to a description of jammed packings and glasses  based on 4 exponents, with three scaling relations between them. We have in particular $\kappa=2/(1+\gamma)$, both sides of which can be measured independently.

\subsection{Comparison with numerics} To confirm the novel prediction that $D(\omega)$ is not flat but scales with frequency as jamming is approached from the hard sphere side, we perform numerical simulations of a hard-sphere glass in $d=2$, at pressure $p=10^{12} k_B T$ and volume fraction $\phi \approx 0.83$ (details are in the SI). The density of states $D(\omega)$ can be computed by identifying a contact network via time averaging as done in \cite{Brito06,Brito09}. Our result for the largest pressure is shown in Fig. \ref{fig:Domega3}, confirming the presence of a weak divergence of $D(\omega)$ with frequency. The exponent appears close to that predicted by Eq.(\ref{4}), but larger simulations are needed, preferably in $d=3$ to avoid logarithmic corrections. We note that this prediction could be tested in colloidal systems using static pair correlation to extract ${\cal M}$ and $D(\omega)$ \cite{Ghosh10,Chen10,Ikeda13,Kaya10,Mari09}. 

\section{Comparison with Replica Theory in {\small $d=\infty$}:} 
\begin{figure}[t] 
\centering
\begin{tikzpicture}[scale=1]
\clip (0,-0.1) rectangle (5.8,4.5);
\includegraphics[viewport=30 10 420 320,width=0.30\textwidth]{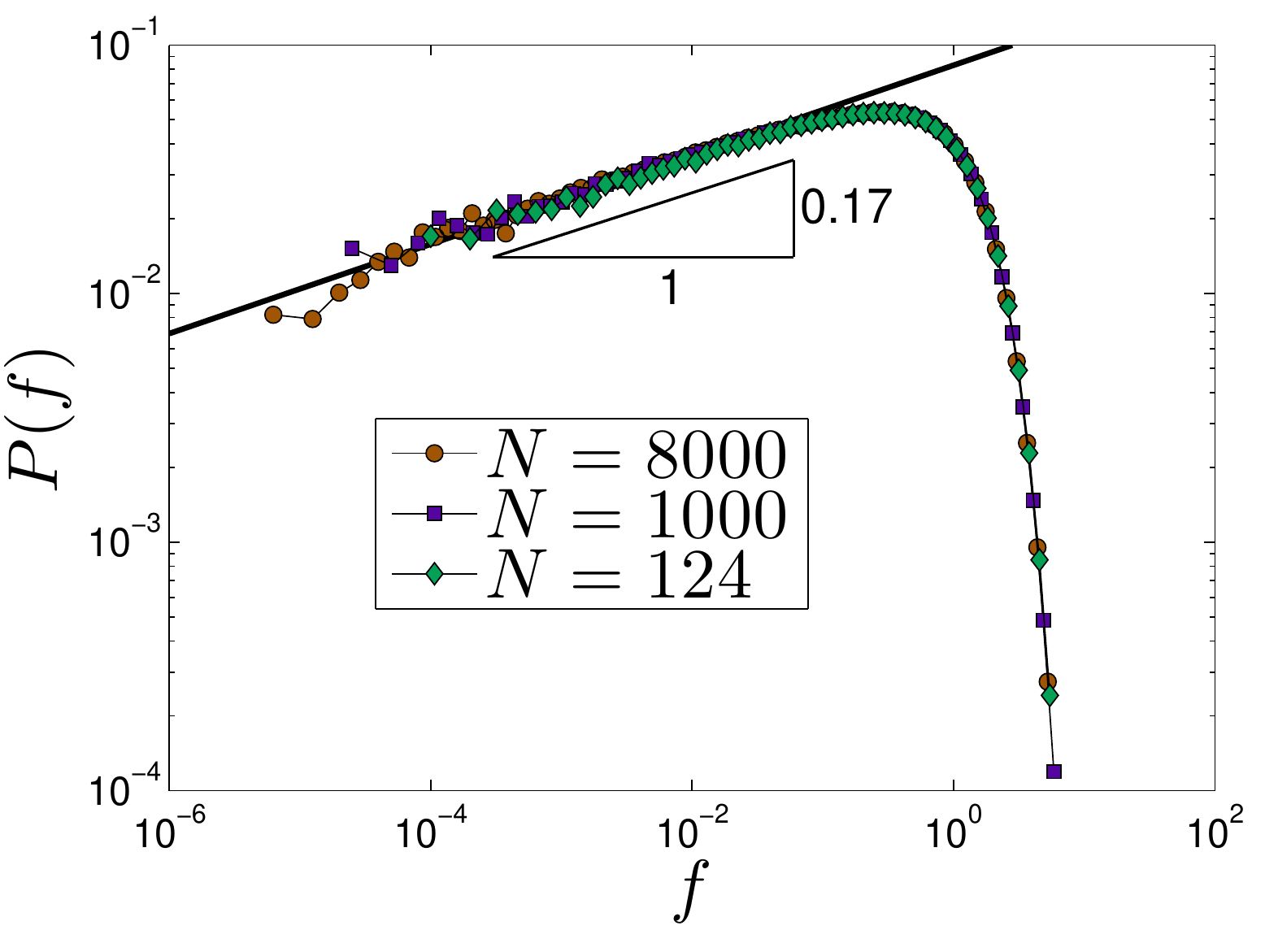}
\end{tikzpicture}
\caption{\label{fig:Pf} Probability distribution of forces $\PP(f)$ for isostatic packings of soft spheres at indicated system sizes, showing that $\PP(f) \sim f^{\theta_f}$ at small $f$, with an exponent $\theta_f=0.17$.}
\end{figure}

A very recent replica computation  \cite{Kurchan12,Kurchan13,Charbonneau13,Charbonneau14} was used to compute exponents in $d=\infty$ to arbitrary precision, and gets  $\gamma=0.41269$, $\kappa=1.41574$ and $\theta_f=0.42311$. These values are consistent with our prediction $\kappa=2/(1+\gamma)$, which appears to be exactly satisfied. 
However, the numerical value we found previously \cite{Lerner12,Lerner13} for $\theta_f\approx 0.17$  in two and three dimensions differs from the replica computation at $d=\infty$. It was argued based on numerics \cite{Charbonneau12} that exponents weakly depend on spatial dimensions up to $d=10$, leading to the suggestion that dimension does not play a role. The same work also reported that $\theta_f$ depends somewhat on system preparation. To check that our value of $\theta_f$ is not due to the specific methods we used (in \cite{Lerner12} results were obtained in two dimensions by shear-jamming hard disks, while in \cite{Lerner13}  hard spheres were compressed in an over-damped medium), we repeat the measurement of force distribution by decompressing soft spheres as done in \cite{Charbonneau12}, but with much higher statistics for the dimension considered. Figure \ref{fig:Pf} shows $\PP(f)$ in 3 dimensions, and again we find $\theta_f = 0.17 \pm 0.02$ (details appear in the SI). Our results therefore support that system preparation does not affect the exponent $\theta_f$, and that its value is indeed about $0.17$ for the bi-disperse system used. Note that for mono-disperse packings in 3 dimensions our numerics suggest a slightly larger exponent $\theta_f \approx 0.23$ as shown in SI. 



The value for $\theta_f$ in $d=\infty$ is therefore distinct from its value in $d=2,3$, and our relation (\ref{10}) is not satisfied in $d=\infty$. This is puzzling, because $\gamma$ appears to be independent of dimension \cite{Charbonneau12,Lerner13}. To resolve this dilemma, note that Eq.(\ref{11}) is also exactly satisfied by the $d=\infty$ result if $\theta_f=\theta_e$. This suggests a simple reconciliation: if it is assumed that localized excitations do not exist for $d=\infty$, then $\theta_f=\theta_e$, and one is left with 3 exponents constrained by two scaling relations: Eq.(\ref{11})  (where $\theta_f=\theta_e$), and Eq.(\ref{7}), both exactly satisfied in the replica calculation. The scaling description we propose based on the marginality of real space excitations (both linear and non-linear) is thus fully consistent with the replica calculation, as these two scaling relations are satisfied.

The fact that localized excitations appear to be absent in large dimension seems plausible, as their existence depends on the presence of local arrangements of particles that are very soft, illustrated in Figure \ref{figlocal}, which may become unlikely when each particle shares many contacts.  This situation may be similar to the behavior of `rattlers', particles which are trapped in a packing but do not contribute to mechanical stability. The fraction of rattlers is observed to decay {\it exponentially} with $d$ \cite{Charbonneau12}, so that in large dimension, it is extremely rare to find a gap that is large enough to hold a particle. The same exponential decay may occur for localized excitations. 

\subsection{Conclusion} We have shown that the stability of hard spheres glasses is affected by  heterogeneity in contact strengths. Our new numerics on the force distribution exponent $\theta_f$, together with the marginal stability relations described above, support that the key exponent $\theta_e \approx 0.41$ in $d=2$ and $d=3$, independent of system preparation. This yields specific predictions for the exponents (\ref{exp}):
\eq{ \label{exp2}
a = 0.17, \;\; b = 0.21, \;\; \kappa = 1.41
} 
If localized excitations are absent in large dimension, then our results are fully consistent with the replica theory; in this case the exponent $\theta_e = 0.42311..$ and the exponents (\ref{exp2}) may change in their final digit.

Our scaling predictions on $D(\omega)$, $\langle \delta R^2 \rangle$, and $\mu$, Eqs. (\ref{6bis}) and (\ref{7}), may be tested experimentally in colloidal systems. From the covariance matrix of particle displacements, ${\cal C}_{ij} \equiv \langle \delta \vec{R}_i \; \delta \vec{R}_j \rangle$, one may define a stiffness matrix ${\cal M}_{ij} \equiv (m_i k_B T)^{-1} \; {\cal C}_{ij}^{-1}$. The latter corresponds to the stiffness matrix of a system interacting with an effective potential, which for hard spheres is Eq.(\ref{V}). This procedure has been carried out in simulations \cite{Brito06,Brito07,Brito09} and experiments \cite{Ghosh10,Chen10,Kaya10}, confirming the presence of a peak in $D(\omega)$ at low frequency. Our new predictions, Eq.(\ref{exp2}) appear to be accessible experimentally \cite{Bonn14}.



Overall, our approach leads to a description of jamming in finite dimensions based on the marginal stability of three distinct types of excitations, both linear and nonlinear. It remains to be seen if plastic flow under shear and thermally activated process near the glass transition can be expressed in terms of the relaxation of these excitations.

 

\begin{acknowledgments}
We thank the authors of \cite{Charbonneau13} for sharing their preprint and for discussions, and Jie Lin, Le Yan, Gustavo D\"uring, Colm Kelleher, and Marija Vucelja for discussions. MW acknowledges support from NSF CBET Grant 1236378, NSF DMR Grant 1105387, and MRSEC Program of the NSF DMR-0820341 for partial funding.
\end{acknowledgments}

\def\rb{\bm{r}}
\def\qb{\bm{q}}
\def\nb{\bm{n}}
\def\delb{\bm{\hat{\delta}}}
\def\curlyM{\mathcal{M}}
\newcommand{\fbar}{\bar{f}}
\newcommand{\kbar}{\bar{k}}
\newcommand{\OO}{\mathcal{O}}
\newcommand\Gbo{\overline{\Gb}}
\def\Gb{\bm{G}}
\def\kpa{k^\parallel}
\def\kpe{k^\perp}
\def\Gpa{G^\parallel}
\def\Gpe{G^\perp}
\def\p{\partial}
%
%
\section{Supplementary Information}

In this Supplementary Information, we provide (A) details of the effective medium theory discussed in the main text, (B) evidence that including localized modes does not improve the variational argument, (C) information on the hard-sphere numerical simulations, (D) information on the soft-sphere numerical simulations, and (E) evidence that the change in gap distribution at finite $\dz$ does not affect our results. 

\section{A. Effective medium theory}
\renewcommand{\theequation}{A.\arabic{equation}}
\setcounter{equation}{0}

Our effective medium theory (EMT) is an extension of \cite{DeGiuli14}. The difference in the present work is to allow the bond stiffnesses and contact forces to follow nontrivial distributions $\PP(k)$ and $\PP(f)$. For the latter, we consider
\eq{
\PP(f) =C_f f^{\theta} e^{-f/\fbar},
}
($\theta = \theta_f$ in the main text) with $0 \leq \theta< 1$ and contact force law
\eq{
f = k_1 |h|^x,
}
where $h$ is the gap at a contact ($h<0$ for overlap). We are interested in the cases $-1 \leq x<0$ and $x > 1$: the former ($x<0$) corresponds to hard particles, and the latter ($x\geq 1$) corresponds to soft particles. We do not consider cusp-like potentials $0<x<1$. We assume particle diameter $\sigma=1$ so that $k_1$ has units of stiffness. The contact stiffness is $k = -df/dh \propto |h|^{x-1}$. This implies
\eq{ \label{probk}
\PP(k) = C_k k^\alpha e^{-(k/\kbar)^{x/(x-1)}}
}  
with $\alpha = (1+x\theta)/(x-1)$. We have $\alpha >0 $ when $x>1$ and $\alpha < 0$ when $x<0$. The contact strain $e$ is defined by $e \equiv \langle f \rangle / \langle k \rangle$. We take units with $\kbar=1$. 

As in previous work, we model a random elastic network of coordination $z$ by diluting a regular lattice of coordination $z_0$ down to $z$. The stiffness in contact $\alpha$, $k_\alpha$, and the force in the contact, $f_\alpha$ are random variables distributed according to
\eq{
\PP_{EMT}(k_\alpha) & = (1-P) \delta(k_\alpha) + P \; \PP(k_\alpha) \\ 
\PP_{EMT}(f_\alpha) & = (1-P) \delta(f_\alpha) + P \; \PP(f_\alpha) 
}
where $P=z/z_0$ to model random dilution of the lattice. 

In EMT, the elastic behavior of a random material, such as our randomly diluted lattice, is modelled by a regular lattice with effective frequency-dependent stiffnesses; as in \cite{DeGiuli14} we will have a longitudinal stiffness, $\kpa$, and a transverse stiffness $-e\kpe$. Writing $\overline{\; \cdot \; }$ for disorder average, the EMT equations are, from \cite{DeGiuli14} \footnote{Here we correct several typos in that work},
\eq{
0 = \overline{ \frac{\kpa-k_\alpha}{1-(\kpa-k_\alpha) G^\parallel} } = \overline{ \frac{e \kpe-f_\alpha}{1+(e \kpe-f_\alpha) G^\perp} },
}
where $\Gpa$ and $\Gpe$ are related to the Green's function $\Gb(\omega) = \big(\curlyM - m\omega^2\big)^{-1}$ by
\eq{
G^\parallel & = \nb_\alpha \cdot \langle \alpha | \Gb | \alpha \rangle \cdot \nb_\alpha \\
G^\perp & = \frac{1}{d-1} \left[ \tr( \langle \alpha | \Gb | \alpha \rangle) - G^\parallel \right],
}
with $\langle \alpha | \equiv \langle i | - \langle j|$. In the present case this leads to
\eq{ \label{em1}
0 & = \frac{(1-P) \kpa}{1-\kpa G^\parallel} + \frac{P C_k}{\Gpa} \left[ -\frac{1}{C_k} + \frac{\beta}{\Gpa} \int_{0}^\infty df \; \frac{f^{\theta} e^{-f}}{c+f^{\beta}} \right]  \\ 
0 & = \frac{(1-P) e \kpe }{1+e \kpe G^\perp} + \frac{P C_f}{\Gpe} \left[\frac{1}{C_f} - \frac{\fbar^{\theta}}{G^\perp} \int_0^\infty df \; \frac{f^{\theta} e^{-f}}{c_2 - f}  \right], \label{em2}
}
with $\beta=1-1/x$, $c=(1-\kpa \Gpa)/\Gpa$, and $c_2 = (1+e\kpe \Gpe)/(\fbar \Gpe)$. These equations need to be supplemented with an equation for $\Gb$. As in \cite{DeGiuli14}, we consider a simplified continuum-like Green's function with a single elastic modulus, and whose isotropy has been restored. This is 
\def\etil{\tilde{e}}
\eq{ \label{G1}
\Gb(\rb,\omega) = \frac{z_0}{d} \delb \int_{BZ} \frac{d^d q}{(2\pi)^d} \frac{e^{i \qb \cdot \rb}}{(\kpa-\etil \kpe) q^2 - m \omega^2},
}
where $BZ = \{ \qb: \; |\qb| < \Lambda  \}$ is an approximate first Brillouin zone, $\etil=(d-1)e$, and $\delb$ is the identity tensor. Isotropy of $\Gb$ implies an identity
\eq{ \label{G2}
G^\parallel = G^\perp & = \frac{2d}{z_0} \frac{1}{\kpa-\etil \kpe} \left( 1 + \frac{m \omega^2}{d} \tr(\Gb(0,\omega)) \right).
}
We solve equations \eqref{em1}, \eqref{em2}, \eqref{G1}, and \eqref{G2} in the limit $e \ll 1$ and $\dz = z - z_c \ll 1$, for $\omega \ll 1$ (we now take $m=1$). Based on previous results \cite{DeGiuli14}, we expect $|c| \ll 1$ and $|c_2| \gg 1$ (which can be checked \textit{a posteriori}), which allows an expansion
\eq{
\int_{0}^\infty df \; \frac{f^{\theta} e^{-f}}{c+f^{\beta}} & = \begin{cases} - \frac{c^\alpha \pi}{\beta \sin(\pi \alpha)} + \ldots & \mbox{if \;} -1 < \alpha < 0 \\
\Gamma_{\alpha \beta} +\ldots & \mbox{if \;} \alpha > 0, \end{cases}
}
with $\Gamma_t= \int_0^\infty x^{t-1} e^{-x} dx$. From this result it can deduced that for $\alpha > 0$, the previous results of \cite{DeGiuli14} are obtained, up to prefactors which depend on $\theta$ and $x$. Therefore, for soft particles with $\alpha \geq 0$, the scalings of \cite{DeGiuli14} are unchanged by stiffness heterogeneity, and henceforth we only consider the case $\alpha < 0$, corresponding to an abundance of weak springs, as discussed in the main text. The other integral is found similarly
\eq{
\int_0^\infty df \; \frac{f^{\theta} e^{-f}}{c_2 - f} = \frac{\Gamma_{\theta+1}}{c_2} + \frac{\Gamma_{\theta+2}}{c_2^2} +  \OO(1/c_2^3) 
}
The leading order EMT equations are then
\eq{ \label{eqns}
0 & = \kpa \Gpa - P + \frac{(1-\kpa \Gpa)^{\alpha+1} P}{{\Gpa}^{\alpha+1} \Gamma_{\theta+ 1}} \frac{\pi}{\beta \sin(\pi |\alpha|)} \\
0 & = e \kpe \Gpe - \frac{P \fbar \Gpe (\theta+ 1)}{1+e\kpe \Gpe}  \label{eqns2}
}
Assuming $\omega \ll \sqrt{\kpa/m}$, it can be checked that
\eq{
\frac{1}{d} \mbox{tr}[\Gb(0,\omega)] = \frac{A_1}{\kpa-\etil\kpe} + \ldots
}
with
\eq{
A_1 = \frac{z_0}{d} \frac{2 \pi^{d/2}}{\Gamma_{d/2} (2\pi)^d} \begin{cases} \frac{\Lambda^{d-2}}{d-2} & \mbox{if } d \geq 3 \\ \half \log(1/\dz) & \mbox{if } d=2 \end{cases}
}
The above equations can be solved for $\dz \ll 1$ following the procedure in \cite{DeGiuli14}: we let
\eq{
\kpa & \sim \dz^\xi, \quad e \kpe \sim \dz^\eta, \quad e = e_c e' \\
e_c & \sim \dz^\chi, \quad \omega \sim \dz^\zeta
}
and balance terms in the above equations. Note that $e$ and $\dz$ are independent parameters: in an elastic network they can be controlled independently. Here $e_c$ is the critical contact strain at elastic instability \cite{DeGiuli14}. One finds 
\eq{
\xi & = 1/(\alpha+1), \\
\chi = \eta = 2 \zeta & = 1+\xi = \frac{\alpha+2}{\alpha+1}
} 
reproducing the scalings in the main text. To leading order, the transverse stiffness is
\eq{
\kpe = \frac{2d}{z_0} \frac{(\theta+1) \Gamma_{\theta+2-1/x}}{\Gamma_{\theta+2}}
}
while the leading order equation for $\kpa$ is 
\eq{ \label{eqn!}
0 = 2d A_1 \omega^2 - \kpa \dz + \etil \kpe z_c + {\kpa}^{\alpha+2} c_3  
}
with $c_3 = \pi z_0 (1-2d/z_0)^{\alpha+1} (z_0/(2d))^{\alpha}/(\Gamma_{\theta+1} \sin(\pi|\alpha|)).$ This is a transcendental equation for $\kpa$ that does not have an analytic solution. However, we can determine some of its key properties.

We expect an onset frequency $\omega_0$ where the density of states $D(\omega)$ grows from 0. This requires that at $\omega_0$, $|d\kpa/d\omega|=\infty$, giving
\eq{
\omega_0 = \omega^* \sqrt{1-e/e_c}
}
with $\omega^* = \dz^\zeta \sqrt{c_4/(z_c A_1)}$, $e_c =  \dz^{2\zeta} c_4/(2d(d-1)\kpe)$, and $c_4 = (c_3(\alpha+2))^{-1/(\alpha+1)}/\eta$. The onset frequency $\omega_0$ vanishes at elastic instability $e=e_c$. 

Below $\omega_0$, $\kpa~\approx~\kpa_*~\equiv~\kpa(\omega_0)~=~(\dz/(c_3(\alpha+2)))^{1/(\alpha+1)}$. For $\omega \gg \omega_0$, we find instead $\kpa \approx \omega ^{2/(\alpha+2)} (-2d A_1/c_3)^{1/(\alpha+2)}$. Combining these gives the approximate solution 
\eq{
\kpa \approx \kpa_* + \omega ^{2/(\alpha+2)} (-2d A_1/c_3)^{1/(\alpha+2)}.
} 
The density of states is determined by 
\eq{
D(\omega) & = \frac{z_0}{\pi \omega} \mbox{Im}[(\kpa-\etil\kpe)\Gpa] \\
& =  \frac{2dA_1}{\pi} \omega \mbox{Im}[1/(\kpa-\etil\kpe)] + \ldots
}
which readily gives
\eq{
D(\omega) \sim \begin{cases} 0 & \mbox{if } \omega < \omega_0 \\
\omega^{1+2/(\alpha+2)} \dz^{-2/(\alpha+1)} & \mbox{if } \omega_0 < \omega < \omega^* \\
\omega^{1-2/(\alpha+2)} & \mbox{if } \omega > \omega^* \end{cases}
}
Debye behavior is absent below $\omega_0$, but would appear to next order in $\dz$ \cite{DeGiuli14}.

For a marginally stable material, $1-e/e_c \ll 1$ so that $\omega_0=0$. Hard spheres correspond to $x=-1$ and $k_1=k_B T \sim 1$ in our units. The predicted behavior in this case is shown in Fig. 2 in the main text, for $d=2$, corresponding to hard disks. Note that for hard disks, assuming $\theta\approx 0.41$, we have $\alpha =-0.30$, $1+2/(\alpha+2)=2.17$, and $1-2/(\alpha+2)=-0.17$.

The shear modulus is $\mu=\kpa(\omega=0)$. When $e=0$, we find
\eq{
\mu(e=0) = \left(\frac{\dz}{c_3}\right)^{1/(\alpha+1)},
}
while when $e=e_c$, $\mu(e=e_c)=\kpa_*$, so that $\mu$ is smaller by a factor of
\eq{
\frac{\mu(e=0)}{\mu(e=e_c)} =  (\alpha+2)^{1/(\alpha+1)} 
}
at instability. Note that when $\alpha=0$ we recover the factor $2$ found in earlier theory \cite{DeGiuli14,Yoshino12}.

Finally, as in \cite{DeGiuli14} we can extract the asymptotic behavior of the Green's function for large $r$. To leading order, $\log(\Gb(r,\omega))\sim -r/\ell_s(\omega) + i \omega r/\nu(\omega)$ where $\ell_s(\omega)=-\omega^{-1} |\Delta k|/\mbox{Im}[\sqrt{\Delta k}]$ and $\nu(\omega)=|\Delta k|/\mbox{Re}[\sqrt{\Delta k}]$ are, respectively, the scattering length and sound velocity at frequency $\omega$. Here $\Delta k = \kpa - \etil \kpe$. The former behaves as
\eq{
\ell_s(\omega) \sim \begin{cases} \infty & \mbox{if } \omega < \omega_0 \\
\omega^{-(4+\alpha)/(2+\alpha)} \dz^{3/(2\alpha+2)} & \mbox{if } \omega_0 < \omega < \omega^* \\
\omega^{-(1+\alpha)/(2+\alpha)} & \mbox{if } \omega > \omega^* \end{cases}
}
while the latter is instead
\eq{
\nu(\omega) \sim \begin{cases} \dz^{1/(2+2\alpha)} & \mbox{if } \omega < \omega^* \\
\omega^{1/(2+\alpha)} & \mbox{if } \omega > \omega^* \end{cases}.
}
We expect that a Rayleigh scattering regime would appear for $\omega < \omega_0$, at the next order in $\dz$. From these results we note particularly that $\ell_s(\omega^*) \sim \dz^{-1/2}$.

The above results give the leading order behavior when $\dz \ll 1$. In $d=2$, the next terms are smaller only by a factor $\sim 1/(\log 1/\dz)$, leading to significant corrections. Therefore, the plot Fig.(\ref{fig:Domega1}) uses the full form of the Green's function, i.e., 
\eq{
\frac{1}{d} \mbox{tr}[\Gb(0,\omega)]|_{d=2} = \frac{A_1}{\Delta k} \left[ \log(\Delta k \Lambda^2 - \omega^2) - \log(-i 0^+ - \omega^2) \right], 
} 
with $A_1= z_0/(8\pi)$. In $d \geq 3$ the next terms are smaller by powers of $\dz$ and this problem does not arise.

\section{B. Localized modes}
\renewcommand{\theequation}{B.\arabic{equation}}
\setcounter{equation}{0}

In the variational argument presented in the main text, we only opened those contacts that led to extended displacements. Here we show that also opening localized contacts, or some fraction of the two populations, does not improve this result.

We use the characterization of small forces described in \cite{Lerner13}. Each contact $\alpha = \langle ij \rangle$ (between particles $i$ and $j$) in an isostatic packing is opened, and the resulting displacement field is measured. Using the fact that the packing is isostatic, the contact force $f_\alpha$ can be written in terms of the resulting displacement field $\vec{\delta R}^{(\alpha)}$. In particular, each force can be written as
\eq{
f_\alpha = f_c \; b_\alpha W_\alpha,
}
where $f_c$ is a typical force, $b_\alpha$ characterizes the strength of far-field displacements relative to the displacements of $i$ and $j$, and $W_\alpha$ characterizes the coupling strength between the displacement $\vec{\delta R}^{(\alpha)}$ and the confining stress (an isotropic pressure in the case considered). In particular, displacements scale as
\eq{ \label{disp}
\vec{\delta R}^{(\alpha)}_{i} \sim \vec{\delta R}^{(\alpha)}_{j} & \sim C, \\
\vec{\delta R}^{(\alpha)}_{k} & \sim b C, \;\;\;\;\; k \neq i,j
}
where $1/C^2 \sim 2 + b^2 N$ is a normalization constant. 

A contact force can be small in two ways: either the far-field displacement field has a small amplitude, $b_\alpha \ll 1$, corresponding to localized modes, or the displacement $\vec{\delta R}^{(\alpha)}$ is weakly coupled to the confining stress, $W_\alpha \ll 1$, corresponding to extended modes. For small values of $b$ and $W$, it was found that 
\eq{
\PP(b) \sim b^{\theta_\ell}, \;\; \PP(W) \sim W^{\theta_e},
} 
and furthermore that $b$ and $W$ are approximately independent. We assume that $\theta_e > \theta_\ell$, as confirmed by numerics, and as implied by marginal stability relations discussed in the main text.

We want to allow, in the variational argument, the possibility of cutting weak links with a certain mix of localized and extended properties. A convenient way to do so is to cut links along the curve 
\eq{
b = W^\eta
}
in $(b,W)$ space, with $0 < \eta < \infty$, so that $f \sim b^{1+1/\eta}$. When $\eta \to 0$, we cut links independently of $b$, corresponding exclusively to extended contacts. When $\eta \to \infty$, we cut links independently of $W$, corresponding exclusively to localized contacts.

Suppose we cut a fraction $q$ of contacts from an isostatic packing. Then the induced $\sim qN$ floppy modes will have displacements scaling as in Eq.(\ref{disp}), but where $i$ and $j$ correspond to any of the particles adjacent to the cut contacts. Modifying accordingly the normalization constant $C$, the energy of a typical mode in the original uncut system will be
\eq{
E \sim \frac{qN k_0}{qN + b^2 N} = \frac{q k_0}{q + b^2},
}
where the stiffness $k_0$ is determined by $q= \int_0^{k_0} \PP(k) dk$. This corresponds to a force $f_0$ with $q= \int_0^{f_0} \PP(f) df$. It follows after some algebra that $q \sim f_0^{1 + (\eta \theta_\ell + \theta_e)/(1+\eta)}$ and $b^2 \sim q^{2\eta/(1+\eta+\eta\theta_\ell+\theta_e)}$. Fixing $q \ll 1$, the best bound is obtained by minimizing the energy $E(q)$, since this corresponds to the smallest frequency for a given amount of cut contacts, and therefore the largest $D(\omega(q)) \gtrsim q/\omega(q)$. There are two cases:

\subsection{Case (i). Predominantly localized contacts $q \gg b^2$:} The condition $q \gg b^2$ requires $\eta > \eta_0$ with $\eta_0 = (1+\theta_e)/(1-\theta_\ell)$. In this case $E \sim k_0 \sim f_0^2 \sim q^{g_1(\eta)}$ with $g_1(\eta) = 2(1+\eta)/(\eta+\eta \theta_\ell + 1+\theta_e)$. It can be checked that $g_1'(\eta)>0$ for all $\eta$, so that the energy is minimized at the largest value of $\eta$, i.e., $\eta \to \infty$. In this case
\eq{ \label{E}
E_{\eta \to \infty} \sim q^{2/(1+\theta_\ell)}
} 
\subsection{Case (ii). Predominantly extended contacts $q \ll b^2$:} The condition $q \ll b^2$ requires $\eta < \eta_0$. In this case $E \sim q k_0/b^2 \sim q^{g_2(\eta)}$ with
\eq{
g_2(\eta) = 1 + \frac{2}{1+\eta+\eta \theta_\ell + \theta_e}.
}
It can be checked that $g_2'(\eta)<0$ for all $\eta$, so that the energy is minimized at the smallest value of $\eta$, i.e., $\eta \to 0$. In this case
\eq{
E_{\eta \to 0} \sim q^{1 + \frac{2}{1+\theta_e}}
}
Now we note that $1 + 2/(1+\theta_e) > 1+2/(1+1) = 2$, and $2/(1+\theta_\ell)<2$. This implies that $E_{\eta \to 0} \ll E_{\eta \to \infty}$ and therefore the smallest energy is attained when choosing only the extended contacts.

\section{C. Hard-sphere numerical simulations}
\renewcommand{\theequation}{C.\arabic{equation}}
\setcounter{equation}{0}

We simulate hard disks using an event-driven molecular dynamics code \cite{Allen89}, in which particles are in free flight until they collide elastically. The system is 50:50 bidisperse, with size ratio 1.4.  We take units with small diameter $\sigma_1=1$, mass $m=1$ (the same for both species), and $k_B T=1$, so that time is measured in units of $\sqrt{m \sigma_1/(k_B T)}$. To generate very large packings, we start with random configurations at very low density and use the Lubachevsky-ÐStillinger algorithm, in which particles are inflated \cite{Donev05a}. The particle inflation rate $\Gamma$ varies with pressure $p$ as $\Gamma = 10^{-3}$ up to $p = 10^2$ and $\Gamma= 10^{-5}$ up to $p = 10^{12}$. At $p = 10^{12}$  the packing fraction is distributed around $\phi_c \approx 0.83$. This protocol generates isostatic packings at $\phi_c$, as was explicitly checked in all the packings used. To obtain configurations at $\phi<\phi_c$, particles are then deflated by a relative amount $\epsilon$, and assigned random velocities. Note that $\half N z k_B T = p (V-V_c) \approx p (\phi_c-\phi) N 3\pi/(2\phi_c^2)$ so that $p (\phi_c-\phi) \approx 0.29 k_B T$.  

To measure the vibrational spectrum of hard disks, it is necessary to define a contact force network within an interval of time $\tau$ \cite{Brito06,Brito09}. Two particles are said to be in contact if they collide with each other during $\tau$. In this same interval, we define $h_{ij}$ as the average gap between two particles and the contact force $f_{ij}$ as the average momentum 
they exchange per unit of time. We can then define an effective potential $V_{eff} = -k_B T \log h_{ij}$ \cite{Brito06,Brito09}, which allows a computation of the dynamical matrix $\cal M$. 
In this work we choose $\tau=1000N$ collisions and $N = 4096$ particles. For larger $\tau$, the vibrational spectrum does not change in the frequency range shown.

\section{D. Soft-sphere numerical simulations}
\renewcommand{\theequation}{D.\arabic{equation}}
\setcounter{equation}{0}

We prepare three-dimensional isostatic packings of bi-disperse soft-spheres, of which half are large and half are small, with the ratio of their respective radii set to 1.4. With $\rho_i$ denoting the radius of the $i^{\rm th}$ particle, and $r_{ij}$ denoting the pairwise distance between the centers of particles $i$ and $j$, the pairwise potential reads $\phi(r_{ij}) = \frac{k}{2}( r_{ij} - (\rho_i + \rho_j) )^2$, where $k$ is the stiffness. We generate isostatic packings by performing a fast quench of a random configuration using the FIRE algorithm \cite{Bitzek06} and applying compressive or expansive strains followed by additional quenches to obtain the target coordination of $z_c = 6$. We choose the stopping condition of the quenches to be $||\vec{F}^{\mbox{\tiny max}}||/\langle f \rangle < 10^{-8}$, where $||\vec{F}^{\mbox{\tiny max}}||$ is the magnitude of the maximum (over all particles in a packing) of the net force, and $\langle f \rangle$ is the mean contact force. We note that for our largest systems of $N=8000$ particles, the isostatic point occurs at dimensionless pressures of the order $10^{-9}$ or smaller; equilibrating packings mechanically at such pressures requires quad floating point precision numerics. 

\begin{figure}[t] 
\includegraphics[width=0.45\textwidth]{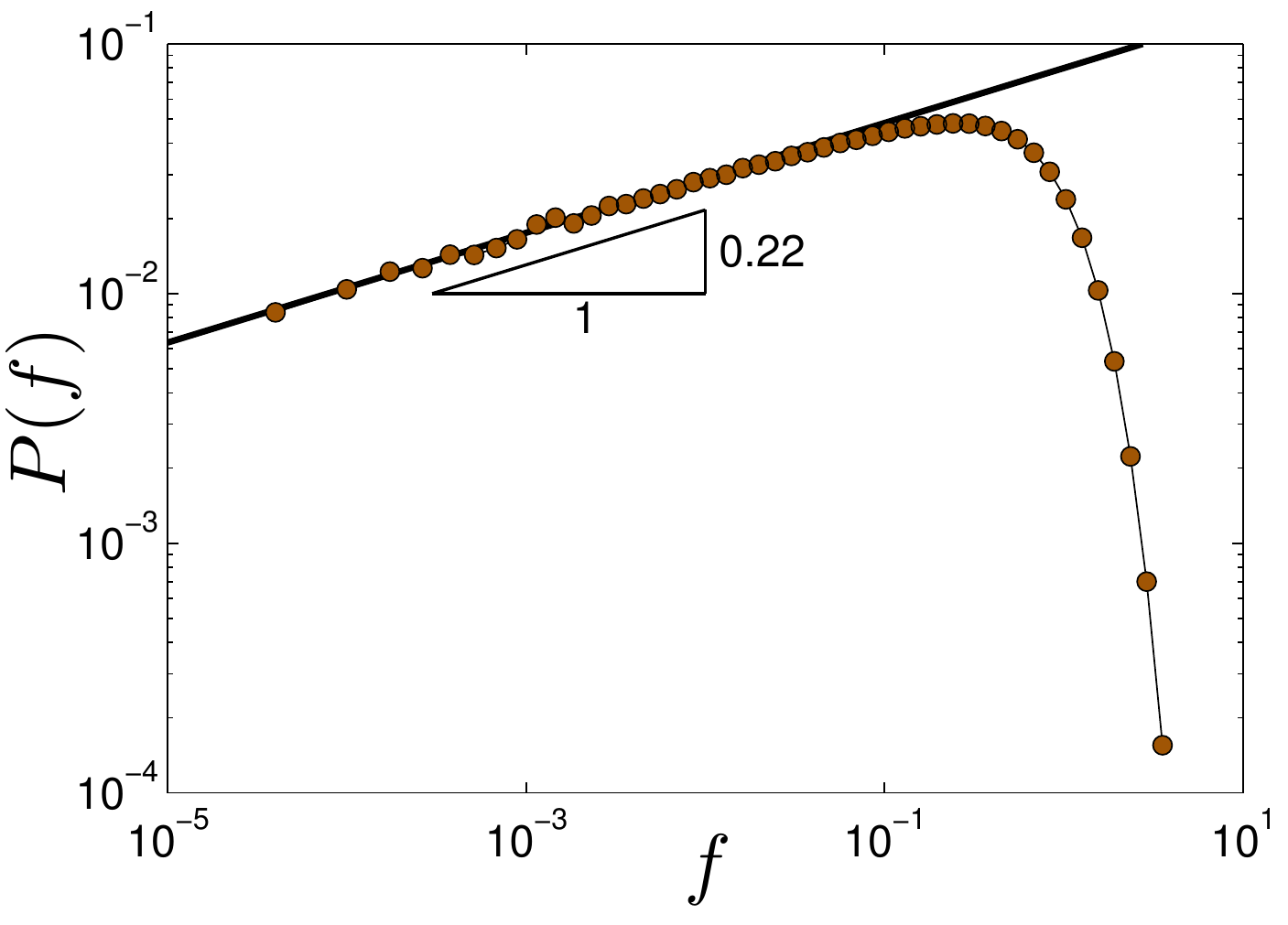}
\caption{\label{fig:monodisperse_forces} Distribution of rescaled contact forces $P(f)$ measured in isostatic packings of $N=4000$ monodisperse harmonic spheres in three dimensions.}
\end{figure}

For the sake of comparison, we have also prepared an ensemble of mono-disperse isostatic packings of $N=4000$, using the same procedure described above. The associated distribution of contact forces $P(f)$ is presented in Fig.~\ref{fig:monodisperse_forces}. We find $P(f) \sim f^{\theta_f}$ with $\theta_f\approx0.22$, which is slightly larger than what we observe in the bi-disperse isostatic packings, suggesting that $\theta_f$ might not be universal.

\section{E. Effect of change of stiffness distribution with $\phi$}
\renewcommand{\theequation}{E.\arabic{equation}}
\setcounter{equation}{0}

In the main text and in the EMT described above, we have assumed that the shape of the distribution of stiffnesses, $\PP(k)$, is independent of $\dz$ and $e$. For hard spheres, we have $f = k_B T/ h$ and $k = k_B T/ h^2$, where $ h $ is the average gap between particles, given that they share a contact (in the sense of \cite{Brito09}). The main effect of changing $\phi$ is to rescale the characteristic stiffness $k_0$, which is included in our approach. However as discussed in \cite{Charbonneau12} one expects the rescaled distribution of gaps (and therefore of stiffnesses) to evolve as $\phi$ departs from $\phi_c$ at weak forces. Here we argue that this evolution, and the presence of additional contacts at large distance and small force,  does not alter our prediction on $\kappa$. For simplicity we shall consider that all particles at distance $h\lesssim 1$ share a contact (a scenario presumably much worse than what occurs in packings where contacts are plausibly not made as soon as $h \gg h_\dagger$ defined below). 
%
We let $k_B T = 1$. 

The hard-sphere gap distribution $g(h)$ has 2 scaling regimes (denoted Ib and IIIb in \cite{Charbonneau12}), and an intermediate matching regime (denoted IIb in \cite{Charbonneau12}). In the first scaling regime, corresponding to gaps that become contacts in the limit $p \to \infty$, we have
\eq{
g(h) \sim p (hp)^{-2-\theta_f} \qquad \mbox{if } h\sim p^{-1}
}
In the second scaling regime, corresponding to gaps that are small, but not zero, as $p \to \infty$, we have
\eq{
g(h) \sim h^{-\gamma} \qquad \mbox{if } h \sim 1
}
In \cite{Charbonneau12}, these forms are shown to match smoothly in an intermediate regime $h\sim p^{-\mu}$ with $\mu=(1+\theta_f)/(2+\theta_f-\gamma)$. Here it will be sufficient to eliminate this intermediate regime by joining the two primary distributions at an intermediate gap size $h_\dagger \sim p^{-\mu}$. We also truncate $g(h)$ at microscopic and macroscopic gap sizes $\delta \ll p^{-1}$ and $h_L \sim 1$. 
We therefore consider
\eq{
g(h) \sim \begin{cases} & p (hp)^{-2-\theta_f} \qquad \mbox{if } \delta < h < h_\dagger   \\ 
& C_2 h^{-\gamma} \qquad\quad \mbox{if } h_\dagger < h < h_L \\ \end{cases}
}
This implies
\eq{
\PP(k) = \frac{C_1}{k^{3/2}} \begin{cases} p (p/\sqrt{k})^{-2-\theta_f} \qquad & \mbox{if } k_\dagger < k < 1/\delta^2 \\
C_2 k^{\gamma/2} \qquad\quad & \mbox{if } k_L < k < k_\dagger \end{cases} 
}
where $k_L = 1/h_L^2$ and $k_\dagger = 1/h_\dagger^2$. The constants $C_1$ and $C_2$ in these expressions are set by requiring that $\PP(k)$ is normalized, and the distribution is continuous at $h_{\dagger}$. This implies $C_2 = p^{-1-\theta_f} k_\dagger^{(2+\theta_f-\alpha)/2}$. 

As discussed in the main text, we consider only the subset of `extended' contacts, in effect replacing $\theta_f$ by $\theta_e$ in this expression. Then since $-3/2 + 1 + \theta_e/2 = \alpha$ and the cutoff $1/\delta^2$ plays the same role as an exponential cutoff (as in the EMT, Eq. \eqref{probk}), this distribution differs from what is considered in the main text by the ultra-weak force regime $k < k_\dagger$. To show that the presence of this regime does not affect our results, we estimate its relative contribution to the energy in a typical mode, $R$, as
\eq{
R & = \frac{\int_{k_L}^{k_\dagger}  dk \; k \; \PP(k)}{\int_{k_\dagger}^{1/\delta^2} dk \; k \; \PP(k)} \\
& \sim p^{1+2\theta_e} \delta^{4+2\alpha} 
}
We can let $\delta \sim p^{-\nu}$ with $\nu \geq 1$, which implies $R \lesssim p^{\theta_e-2}$. This goes to zero as $p\to \infty$, so to leading order the ultra-weak springs contain only an infinitesimal fraction of energy, and will not affect our results.
 
We note that our prediction for $\dz(p)$ discussed in the main text is satisfied in the numerics of Ref. \cite{Charbonneau12}, if the contact network is assumed to consist of those particles whose gap is smaller than $h \sim h_\dagger$. Since our estimate of $R$ assumes contacts are made for $h \lesssim 1$, we expect that $R$ is in fact an upper bound on the contribution of the ultra-weak forces.



\bibliography{../../bib/Wyartbibnew}

\begin{thebibliography}{10}

\bibitem{Goldstein69}
Goldstein M (1969) {Viscous Liquids and the Glass Transition: A Potential
  Energy Barrier Picture }.
\newblock {\em J. Chem. Phys.} 51:3728.

\bibitem{Kirkpatrick89}
Kirkpatrick TR, Thirumalai D, Wolynes PG (1989) Scaling concepts for the
  dynamics of viscous liquids near an ideal glassy state.
\newblock {\em Phys. Rev. A} 40:1045--1054.

\bibitem{Berthier11b}
Berthier L, Biroli G (2011) Theoretical perspective on the glass transition and
  amorphous materials.
\newblock {\em Reviews of Modern Physics} 83:587.

\bibitem{Singh85}
Singh Y, Stoessel J, Wolynes P (1985) Hard-sphere glass and the
  density-functional theory of aperiodic crystals.
\newblock {\em Physical review letters} 54:1059.

\bibitem{Parisi10}
Parisi G, Zamponi F (2010) Mean-field theory of hard sphere glasses and
  jamming.
\newblock {\em Reviews of Modern Physics} 82:789.

\bibitem{Brito06}
Brito C, Wyart M (2006) On the rigidity of a hard-sphere glass near random
  close packing.
\newblock {\em EPL (Europhysics Letters)} 76:149.

\bibitem{Brito09}
Brito C, Wyart M (2009) Geometric interpretation of previtrification in hard
  sphere liquids.
\newblock {\em The Journal of Chemical Physics} 131:024504.

\bibitem{Liu10}
Liu AJ, Nagel SR, van Saarloos W, Wyart M (2010) {\em The jamming scenario: an
  introduction and outlook}, eds.{} L.Berthier, Biroli G, Bouchaud J, Cipeletti
  L, van Saarloos W.
\newblock (Oxford University Press, Oxford).

\bibitem{Hecke10}
van Hecke M (2010) Jamming of soft particles: geometry, mechanics, scaling and
  isostaticity.
\newblock {\em Journal of Physics: Condensed Matter} 22:033101.

\bibitem{Wyart05b}
Wyart M (2005) On the rigidity of amorphous solids.
\newblock {\em Annales de Phys} 30 (3):1.

\bibitem{Maxwell64}
Maxwell J (1864) On the calculation of the equilibrium and stiffness of frames.
\newblock {\em Philos. Mag.} 27:294--299.

\bibitem{Ikeda13}
Ikeda A, Berthier L, Biroli G (2013) Dynamic criticality at the jamming
  transition.
\newblock {\em The Journal of Chemical Physics} 138:12A507.

\bibitem{Ghosh10}
Ghosh A, Chikkadi VK, Schall P, Kurchan J, Bonn D (2010) Density of states of
  colloidal glasses.
\newblock {\em Physical review letters} 104:248305.

\bibitem{Chen10}
Chen K et~al. (2010) Low-frequency vibrations of soft colloidal glasses.
\newblock {\em Physical review letters} 105:025501.

\bibitem{Kaya10}
Kaya D, Green N, Maloney C, Islam M (2010) Normal modes and density of states
  of disordered colloidal solids.
\newblock {\em Science} 329:656--658.

\bibitem{Mari09}
Mari R, Krzakala F, Kurchan J (2009) Jamming versus glass transitions.
\newblock {\em Phys. Rev. Lett.} 103:025701.

\bibitem{Charbonneau13}
Charbonneau P, Kurchan J, Parisi G, Urbani P, Zamponi F (2013) Exact theory of
  dense amorphous hard spheres in high dimension. {III}. the full {RSB}
  solution.
\newblock {\em arXiv preprint arXiv:1310.2549}.

\bibitem{Charbonneau14}
Charbonneau P, Kurchan J, Parisi G, Urbani P, Zamponi F (2014) Fractal free
  energy landscapes in structural glasses.
\newblock {\em Nature communications} 5.

\bibitem{Wyart12}
Wyart M (2012) Marginal stability constrains force and pair distributions at
  random close packing.
\newblock {\em Phys. Rev. Lett.} 109:125502.

\bibitem{Lerner13}
Lerner E, D{\"u}ring G, Wyart M (2013) Simulations of driven overdamped
  frictionless hard spheres.
\newblock {\em Computer Physics Communications} 184:628 -- 637.

\bibitem{Landau60}
Landau LD, Lifshitz E (1960) {\em Theory of Elasticity: Vol. 7 of Course of
  Theoretical Physics}.
\newblock Vol.{}~13, p.~44.

\bibitem{Alexander98}
Alexander S (1998) Amorphous solids: their structure, lattice dynamics and
  elasticity.
\newblock {\em Physics Reports} 296:65--236.

\bibitem{Wyart05}
Wyart M, Nagel S, Witten T (2005) Geometric origin of excess low-frequency
  vibrational modes in weakly connected amorphous solids.
\newblock {\em EPL (Europhysics Letters)} 72:486.

\bibitem{Ohern03}
O'Hern CS, Silbert LE, Liu AJ, Nagel SR (2003) Jamming at zero temperature and
  zero applied stress: The epitome of disorder.
\newblock {\em Phys. Rev. E} 68:011306.

\bibitem{Silbert05}
Silbert LE, Liu AJ, Nagel SR (2005) Vibrations and diverging length scales near
  the unjamming transition.
\newblock {\em Phys.\ Rev.\ Lett.} 95:098301.

\bibitem{Wyart08}
{Wyart} M, {Liang} H, {Kabla} A, {Mahadevan} L (2008) {Elasticity of Floppy and
  Stiff Random Networks}.
\newblock {\em Phys.\ Rev.\ Lett.} 101:215501.

\bibitem{Lerner12}
Lerner E, D\"uring G, Wyart M (2012) Toward a microscopic description of flow
  near the jamming threshold.
\newblock {\em EPL (Europhysics Letters)} 99:58003.

\bibitem{Ellenbroek09}
Ellenbroek WG, Zeravcic Z, van Saarloos W, van Hecke M (2009) Non-affine
  response: Jammed packings vs. spring networks.
\newblock {\em EPL} 87:34004.

\bibitem{Ellenbroek06}
Ellenbroek WG, Somfai E, van Hecke M, van Saarloos W (2006) Critical scaling in
  linear response of frictionless granular packings near jamming.
\newblock {\em Phys. Rev. Lett.} 97:258001.

\bibitem{Garboczi85}
{Garboczi} EJ, {Thorpe} MF (1985) Effective-medium theory of percolation on
  central-force elastic networks .2. further results.
\newblock {\em Phys.\ Rev.\ B} 31:7276.

\bibitem{Schirmacher07}
Schirmacher W, Ruocco G, Scopigno T (2007) Acoustic attenuation in glasses and
  its relation with the boson peak.
\newblock {\em Phys. Rev. Lett.} 98:025501.

\bibitem{Webman81}
Webman I ({1981}) Effective-medium approximation for diffusion on a random
  lattice.
\newblock {\em PRL} {47}:1496--1499.

\bibitem{Wyart10a}
Wyart M (2010) Scaling of phononic transport with connectivity in amorphous
  solids.
\newblock {\em EPL (Europhysics Letters)} 89:64001.

\bibitem{Mao10}
Mao X, Xu N, Lubensky TC (2010) Soft modes and elasticity of nearly isostatic
  lattices: Randomness and dissipation.
\newblock {\em Phys. Rev. Lett.} 104:085504.

\bibitem{DeGiuli14}
DeGiuli E, Laversanne-Finot A, D\"uring GA, Lerner E, Wyart M (2014) Effects of
  coordination and pressure on sound attenuation, boson peak and elasticity in
  amorphous solids.
\newblock {\em Soft Matter} 10:5628--5644.

\bibitem{Sheinman12}
Sheinman M, Broedersz C, MacKintosh F (2012) Nonlinear effective-medium theory
  of disordered spring networks.
\newblock {\em Physical Review E} 85:021801.

\bibitem{HardSphereSI}
(year?).
\newblock See Supplemental Material at XXXX for the effective medium theory,
  and for details about numerical simulations.

\bibitem{Lerner13a}
Lerner E, During G, Wyart M (2013) Low-energy non-linear excitations in sphere
  packings.
\newblock {\em Soft Matter} 9:8252--8263.

\bibitem{Charbonneau12}
Charbonneau P, Corwin EI, Parisi G, Zamponi F (2012) Universal microstructure
  and mechanical stability of jammed packings.
\newblock {\em Physical Review Letters} 109:205501--.

\bibitem{Donev05a}
Donev A, Torquato S, Stillinger FH (2005) Pair correlation function
  characteristics of nearly jammed disordered and ordered hard-sphere packings.
\newblock {\em Phys. Rev. E} 71:011105.

\bibitem{Silbert06}
Silbert LE, Liu AJ, Nagel SR (2006) Structural signatures of the unjamming
  transition at zero temperature.
\newblock {\em Phys.\ Rev.\ E} 73:041304.

\bibitem{Kurchan12}
Kurchan J, Parisi G, Zamponi F (2012) Exact theory of dense amorphous hard
  spheres in high dimension {I}. the free energy.
\newblock {\em Journal of Statistical Mechanics: Theory and Experiment}
  2012:P10012.

\bibitem{Kurchan13}
Kurchan J, Parisi G, Urbani P, Zamponi F (2013) Exact theory of dense amorphous
  hard spheres in high dimension. {II}. the high density regime and the
  {G}ardner transition.
\newblock {\em The Journal of Physical Chemistry B} 117:12979--12994.

\bibitem{Brito07}
Brito C, Wyart M (2007) Heterogeneous dynamics, marginal stability and soft
  modes in hard sphere glasses.
\newblock {\em Journal of Statistical Mechanics: Theory and Experiment}
  2007:L08003.

\bibitem{Bonn14}
Bonn D, Zargar R (2014).
\newblock private communication.

\bibitem{Yoshino12}
Yoshino H (2012) Replica theory of the rigidity of structural glasses.
\newblock {\em The Journal of Chemical Physics} 136:214108.

\bibitem{Allen89}
Allen MP, Tildesley DJ (1989) {\em Computer simulation of liquids}.
\newblock (Oxford university press).

\end{thebibliography}
\bibliographystyle{pnas2011}


\end{article}
\end{document}